\documentclass[aps,pra,twocolumn,superscriptaddress]{revtex4-2}
\usepackage{amsmath}
\usepackage{amssymb}
\usepackage{mathtools}
\usepackage{graphicx}
\usepackage{bm}
\usepackage{braket}
\usepackage{bbm}
\usepackage[dvipsnames]{xcolor}

\newcommand{\tr}{\operatorname{Tr}}
\newcommand{\mse}{\bar{\epsilon}}
\newcommand{\tb}{{\bm{\theta}}}
\newcommand{\nh}{\hat{\bm{n}}}
\newcommand{\Gz}{\hat{\Gamma}_0}
\newcommand{\Go}[1]{\hat{\Gamma}_1^{(#1)}}
\newcommand{\HI}{\mathcal{H}_I}
\newcommand{\HO}{\mathcal{H}_O}
\newcommand{\HA}{\mathcal{H}_A}
\newcommand{\Id}[1]{\mathbbm{1}_{#1}}
\newcommand{\Wrep}{\mathcal{W}}
\newcommand{\seed}{\Xi}
\newcommand{\Lop}{\mathbb{L}}
\newcommand{\su}{\mathrm{SU}(2)}
\newcommand{\so}{\mathrm{SO}(3)}

\begin{document}

\title{Saturating the Bayesian Nagaoka-Hayashi bound within numerical precision for the depolarization SU(2) rotation channel}

\author{Leo Bia}
\email{lclashinbia@arizona.edu}
\affiliation{Wyant College of Optical Sciences, University of Arizona, Tucson, Arizona 85721, USA}

\author{Christos N. Gagatsos}
\affiliation{School of Electrical, Computing, and Software Engineering, University of Arizona, Tucson, Arizona 85721, USA}
\affiliation{Wyant College of Optical Sciences, University of Arizona, Tucson, Arizona 85721, USA}
\affiliation{Program in Applied Mathematics, University of Arizona, Tucson, Arizona 85721, USA}

\date{\today}

\begin{abstract}
The Bayesian Nagaoka--Hayashi (NH) bound is a semidefinite lower bound on the Bayes risk of multiparameter estimation, tighter than the Bayesian symmetric logarithmic derivative (SLD) Cram\'er--Rao bounds, and whether it can be attained is an open problem. We study the single-shot estimation of a qubit rotation, all three parameters at once, under depolarizing noise and a uniform prior, with $k$ parallel uses of the channel. Rotational covariance reduces the joint optimization of probe, measurement, and estimator, and the NH bound itself, to small semidefinite programs, solvable through $k=4$. At every number of uses and noise strength the optimized strategy reaches the NH bound within the numerical precision of the calculation, numerical evidence that the bound is attained for this channel family, while the SLD bound lies strictly below and is never attained. As the noise grows the optimal probe collapses to the tensored Bell state.
\end{abstract}

\maketitle

\section{\label{sec:intro}Introduction}

Quantum resources enable measurement precision beyond classical limits~\cite{Giovannetti2011advances}. For a single parameter the theory is essentially complete. The quantum Cram\'er--Rao bound, built on the symmetric logarithmic derivative (SLD), sets the attainable precision and is saturated by a projective measurement in the SLD eigenbasis~\cite{Helstrom1967,Braunstein1994}. Most sensing tasks are not of this kind. Field sensing, imaging, and device characterization ask for several parameters at once~\cite{Demkowicz2020review,SidhuKok2020}. When the generators of the parameters do not commute, the individually optimal measurements are mutually incompatible, no single measurement attains every single-parameter bound simultaneously, and the attainable error generically exhibits a finite gap to the SLD bound~\cite{Helstrom1974,Ragy2016compatibility}. Quantifying this gap, and finding the strategy that performs best in spite of it, is the central problem of multiparameter quantum estimation.

The local theory, built on locally unbiased estimators at a fixed parameter point, meets the multiparameter problem with a ladder of lower bounds on the attainable error. The SLD bound is joined by the right logarithmic derivative bound~\cite{YuenLax1973} and by the Holevo bound~\cite{Holevo2011book}, the tightest of the three, which stays within a small constant factor of the SLD bound~\cite{Tsang2019thrice} and is attained asymptotically by collective measurements on infinitely many copies of the output~\cite{Kahn2009local,Yamagata2013qlan,Yang2019attaining}. A finite experiment lives far from this guarantee. For separable measurements on finitely many copies, Nagaoka derived a tighter two-parameter bound from the approximate joint diagonalization of noncommuting matrices~\cite{Nagaoka1989,Nagaoka1991}, Hayashi generalized it to any number of observables~\cite{Hayashi1999}, and the resulting Nagaoka--Hayashi (NH) bound was recently extended to arbitrarily many parameters and cast as a semidefinite program (SDP)~\cite{Conlon2021}. Conic programming hierarchies now bound the ultimate precision of separable measurements from both sides and optimize the probe in the same framework~\cite{HayashiOuyang2023,HayashiOuyang2024probe}.

These bounds are frequentist. They constrain estimators that are unbiased near a known operating point, and prior knowledge enters nowhere. When data is scarce and prior information matters, the natural formulation is Bayesian. A prior density is placed on the parameters and the figure of merit is a score averaged over it. For the quadratic score the optimal estimator at fixed measurement is the posterior mean, a structure established by Personick, who initiated Bayesian quantum estimation~\cite{Personick1971}, and placed in the general theory of quantum statistical decisions by Holevo~\cite{Holevo1973}. Quantum versions of the classical Bayesian bound families followed, among them Ziv--Zakai~\cite{TsangZivZakai2012}, Weiss--Weinstein~\cite{LuTsang2016}, and van Trees and Gill--Levit type bounds~\cite{Tsang2020GL}. The van Trees family requires a differentiable prior that vanishes on the boundary of the parameter set~\cite{Tsang2020GL}, while the moment bounds descending from Personick impose no regularity on the prior at all. In that line, Rubio and Dunningham derived a multiparameter Bayesian bound from two prior moments of the output family, and proved it attainable if and only if the Bayesian analogues of the SLD operators commute~\cite{Rubio2020}, with independent derivations in Refs.~\cite{SidhuKok2020,Demkowicz2020review}. Incompatibility therefore limits Bayesian estimation exactly as it limits the local theory. Suzuki carried the NH construction to the Bayesian setting, obtaining a bound computable as an SDP, provably tighter than the Bayesian quantum Cram\'er--Rao bounds, and free of any local unbiasedness condition, and posed its attainability as an open problem~\cite{Suzuki2024}. Most recently, the gap between the attainable single-shot error and the moment bound was proposed as a figure of Bayesian incompatibility, and the attainable error was proved to be at most twice that bound~\cite{Albarelli2026}.

In sensing scenarios the parameters usually enter through a quantum channel rather than a fixed family of states, so the probe carries part of the strategy and must be optimized together with the measurement and the estimator. For group transformations, covariant methods determine the optimal probes and measurements from symmetry, with phase shifts and rotations as paradigmatic examples~\cite{Holevo2011book,Acin2001dynamics,Chiribella2005group, Bagan2004,Demkowicz2011phase}. The object that places the probe, the ancilla, and the measurement on one footing is the quantum tester, introduced in equivalent forms as the quantum strategy of multiround games, the process POVM, and the measuring comb of higher-order quantum operations~\cite{Gutoski2007,Ziman2008,Chiribella2008architecture, ChiribellaPRL180501,Chiribella2009}. Testers turn the design of the single-shot Bayesian protocol, jointly over probe, measurement, and estimator, into an SDP~\cite{Bavaresco2024}, recently extended to $k$ channel uses with a comparison of parallel, sequential, and indefinite-causal-order strategies~\cite{Andre2026}. These strategy optimizations carry no analysis on the bound side. Conversely, the Bayesian NH bound has been studied for families of states with the probe held fixed~\cite{Suzuki2024,Albarelli2026}, and whether any measurement attains it is unknown. How tightly the achievable error, the NH bound, and the SLD bound bracket one another for a concrete channel, at more than one use and under noise, has remained open.

In this work we address these questions for the paradigmatic noncommuting family, the $\su$ rotation channel with all three rotation parameters unknown, under depolarizing noise and a uniform prior. Our contributions are the following. (i) Using the rotational covariance of the problem, we reduce the joint optimization of probe, measurement, and estimator over $k$ parallel channel uses, and the Bayesian NH bound itself, to semidefinite programs whose sizes are set by representation multiplicities, solvable through $k=4$. (ii) At every computed number of uses and noise strength the optimized strategy reaches the NH bound within the numerical precision of the calculation, a numerical observation and not an attainment theorem. (iii) As the depolarization grows the optimal probe collapses to the tensored Bell state. Section~\ref{sec:framework} sets the Bayesian estimation framework, Sec.~\ref{sec:bounds} the SLD and NH bounds, Sec.~\ref{sec:testers} the tester representation and the seesaw, Sec.~\ref{sec:su2} the symmetry reduction, Sec.~\ref{sec:results} the results, and Sec.~\ref{sec:discussion} concludes.

\section{\label{sec:framework}Multiparameter Bayesian estimation}

We consider the estimation of $d$ unknown parameters $\tb=(\theta_1,\ldots,\theta_d)\in\Theta\subseteq\mathbb{R}^d$ that label a family of completely positive and trace-preserving channels $\Lambda_{\tb}\colon\mathcal{L}(\HI)\to\mathcal{L}(\HO)$, where $\HI$ and $\HO$ are the input and output Hilbert spaces of the channel, $\mathcal{L}(\mathcal{H})$ is the space of linear operators on $\mathcal{H}$, and all Hilbert spaces in this work are finite dimensional. The probe may be entangled with an ancilla that the channel does not act on, and we introduce the ancillary Hilbert space $\HA$ here since it is needed in the tester representation of Sec.~\ref{sec:testers}; the ancilla-free case is recovered by setting $\HA\cong\mathbb{C}$. A probe, described by a density operator $\rho_0\in\mathcal{L}(\HI\otimes\HA)$, is sent through the channel, producing the output state $\rho(\tb)=(\Lambda_{\tb}\otimes\mathrm{id}_A)[\rho_0]\in\mathcal{L}(\HO\otimes\HA)$, and a measurement described by a positive operator valued measure (POVM) $\{M_m\}\subset\mathcal{L}(\HO\otimes\HA)$, with $M_m\geq0$ and $\sum_m M_m=\Id{OA}$, returns outcome $m$ with likelihood
\begin{equation}\label{eq:likelihood}
    p(m|\tb)=\tr[M_m\,\rho(\tb)] .
\end{equation}
The outcome set may be continuous, in which case sums over $m$ denote integrals with respect to the outcome measure. Prior knowledge is encoded in a probability density $p(\tb)$ on $\Theta$ with finite second moments. An estimator $\hat{\tb}\colon m\mapsto\hat{\tb}(m)\in\mathbb{R}^{d}$ maps each outcome to a parameter estimate, with components $\hat\theta_i(m)$. The figure of merit is the weighted Bayesian mean square error (MSE),
\begin{equation}\label{eq:mse}
    \mse = \sum_{i=1}^{d} w_i
    \int d\tb\, p(\tb)\sum_m p(m|\tb)\,
    [\hat\theta_i(m) - \theta_i]^2 ,
\end{equation}
where the weights $w_i>0$, normalized to $\sum_i w_i=1$, fix the relative importance of the parameters. For any fixed measurement the optimal estimates are the posterior means $\hat\theta_i(m)=\int d\tb\,p(\tb|m)\,\theta_i$, with $p(\tb|m)\propto p(\tb)\,p(m|\tb)$ the posterior~\cite{Personick1971,Rubio2020}. A choice of probe, measurement, and estimator is a strategy, $\mse$ is its Bayes risk, and this formulation of the problem is that of Rubio and Dunningham~\cite{Rubio2020}. All bounds and optimal estimators in this framework depend on the strategy only through the output family $\rho(\tb)$ and the POVM, so they apply unchanged for any ancilla dimension and for any number of channel uses, both of which only enlarge the Hilbert spaces on which $\rho(\tb)$ and $\{M_m\}$ are defined. The task is to minimize Eq.~(\ref{eq:mse}) jointly over the probe, the measurement, and the estimator.

\section{\label{sec:bounds}Bounds on the Bayes risk}

For a fixed probe, the bounds of this section depend on the output family only through its two prior moments,
\begin{align}
    \Gz &= \int d\tb\, p(\tb)\,\rho(\tb) ,\label{eq:moment0}\\
    \Go{i} &= \int d\tb\, p(\tb)\,\theta_i\,\rho(\tb) ,\label{eq:moment1}
\end{align}
where $\Gz$ is the prior-averaged output state and $\Go{i}$ is the Hermitian first moment associated with $\theta_i$. Lower bounds on the Bayes risk go back to Holevo~\cite{Holevo1973}, and the alternatives are compared in this setting in Ref.~\cite{Albarelli2026}. We use the two most relevant here. The SLD bound is the simplest and sets the incompatibility-free benchmark, and the Nagaoka--Hayashi bound is tighter.

\textit{The SLD bound and its attainability.} The Bayesian symmetric logarithmic derivative (SLD) operators $L_i$ are the unique Hermitian solutions, for $\Gz>0$, of the Lyapunov equations
\begin{equation}\label{eq:lyapunov}
    \tfrac{1}{2}\bigl(\Gz L_i + L_i\Gz\bigr) = \Go{i}.
\end{equation}
The Bayes risk of every strategy obeys the SLD bound
\begin{equation}\label{eq:sld_bound}
    \mse \;\geq\; \mse_\mathrm{SLD} = \mse_\mathrm{prior}
    - \sum_{i=1}^{d} w_i\,\tr\!\bigl(\Gz L_i^2\bigr),
\end{equation}
with $\mse_\mathrm{prior}=\sum_i w_i\int d\tb\,p(\tb)\,\theta_i^2$ the weighted second moment of the prior. The bound is due to Personick for a single parameter~\cite{Personick1971} and to Rubio and Dunningham in the multiparameter case~\cite{Rubio2020}. For a single parameter it is tight, attained by projectively measuring the SLD and reporting the eigenvalue as the estimate~\cite{Personick1971}. For several parameters the bound is attainable if and only if the SLD operators commute pairwise, $[L_i,L_j]=0$, in which case the projective measurement in their common eigenbasis is simultaneously optimal for all parameters~\cite{Rubio2020}. When the $L_i$ do not commute, the individually optimal measurements are incompatible and $\mse_\mathrm{SLD}$ lies strictly below the attainable error~\cite{Helstrom1974,Demkowicz2020review}.

\textit{The Nagaoka--Hayashi bound.} The Bayesian form~\cite{Suzuki2024} of the Nagaoka--Hayashi (NH) bound~\cite{Nagaoka1989,Hayashi1999} is tighter than Eq.~(\ref{eq:sld_bound}). Writing $\Lop_{ij}=\sum_m \hat\theta_i(m)M_m\hat\theta_j(m)$ and $X_i=\sum_m \hat\theta_i(m)M_m$, the risk of any strategy equals
\begin{equation}\label{eq:bayes_exact}
    \mse=\tr[\bar{\mathbb{S}}\,\Lop]-\tr[\bar D X^{\intercal}]
    -\tr[X\bar D^{\intercal}]+\mse_\mathrm{prior},
\end{equation}
with $\bar{\mathbb{S}}=W\otimes\Gz$, $W=\mathrm{diag}(w_i)$, and $\bar D_i=\sum_j W_{ij}\Go{j}$. Positivity of the POVM implies $\Lop\geq XX^{\intercal}$, and relaxing the measurement constraints to this Schur condition, over Hermitian $X_i$ and operator matrices $\Lop$ with Hermitian blocks obeying $\Lop_{ij}=\Lop_{ji}$, gives the convex program
\begin{equation}\label{eq:nh_sdp}
    \mse_\mathrm{NH}=\min_{\Lop\geq XX^{\intercal}}
    \bigl\{\tr[\bar{\mathbb{S}}\Lop]
    -\tr[\bar D X^{\intercal}]-\tr[X\bar D^{\intercal}]\bigr\}
    +\mse_\mathrm{prior}.
\end{equation}
Since the relaxed feasible set contains every measurement-induced point, the bounds and the optimum are ordered as~\cite{Suzuki2024}
\begin{equation}\label{eq:chain}
    \mse_\mathrm{SLD}\leq\mse_\mathrm{NH}\leq\mse^\star
    \leq\mse_\mathrm{prior},
\end{equation}
where $\mse^\star$ is the exact optimum of Eq.~(\ref{eq:mse}) for the given probe.

\section{\label{sec:testers}Tester representation of channel strategies}

For $k$ parallel uses of the channel the probe is a state $\rho_0\in\mathcal{L}\bigl(\HI^{\otimes k}\otimes\HA^{\otimes k}\bigr)$, with one ancilla mode $\HA\cong\HI$ per channel use, the channel acts on the input modes only, giving the output $\rho(\tb)=(\Lambda_{\tb}^{\otimes k}\otimes\Id{A})[\rho_0]$, and the measurement $\{M_m\}\subset\mathcal{L}\bigl(\HO^{\otimes k}\otimes\HA^{\otimes k}\bigr)$ acts jointly on the output and ancilla registers. The matched configuration entails no loss of generality within this class, as the reconstruction below shows, and admits the Choi--Jamio{\l}kowski representation that renders the joint optimization over probes and measurements convex. The natural object that treats the probe, the ancilla, and the measurement on the same footing is the quantum tester, introduced in equivalent forms between 2006 and 2009 as the quantum strategy of multiround quantum games~\cite{Gutoski2007}, as the process POVM~\cite{Ziman2008}, and as the measuring comb of the theory of quantum networks and higher-order operations~\cite{Chiribella2008architecture,Chiribella2009}. The name comes from Ref.~\cite{ChiribellaPRL180501}, which applied testers to the discrimination of memory channels and proved that parallel schemes are optimal among all architectures for the discrimination and estimation of independent unitary channels. Testers were recently applied to single-shot Bayesian channel estimation~\cite{Bavaresco2024} and to multiple uses of the $\su$ rotation channel with a fidelity cost~\cite{Andre2026}, and we follow their formulation.

The $k$-use channel is encoded by its Choi operator
\begin{equation}\label{eq:choi}
    J_{\tb}=\sum_{a,b}\ket{a}\!\bra{b}\otimes
    \Lambda_{\tb}^{\otimes k}\bigl[\ket{a}\!\bra{b}\bigr]
\end{equation}
on $\HI^{\otimes k}\otimes\HO^{\otimes k}$, where $\{\ket{a}\}$ is a product basis of $\HI^{\otimes k}$. The outcome probability of any probe and measurement takes the form $p(m|\tb)=\tr[T_m J_{\tb}]$ with the tester element
\begin{equation}\label{eq:tester_def}
    T_m=\tr_A\!\bigl[(\Id{I}\otimes M_m)
    (\rho_0^{T_I}\otimes\Id{O})\bigr],
\end{equation}
where $(\cdot)^{T_I}$ is the partial transpose on $\HI^{\otimes k}$, so that the entire strategy up to the estimator is absorbed into a single positive operator on $\HI^{\otimes k}\otimes\HO^{\otimes k}$. The set of physical testers is characterized exactly by the two linear conditions~\cite{Chiribella2009,Ziman2008,Bavaresco2024,Andre2026}
\begin{equation}\label{eq:tester_constraints}
    T_m\geq 0\ \ \forall m,
    \qquad
    \sum_m T_m=\sigma\otimes\Id{O},
\end{equation}
with $\sigma=(\tr_A\rho_0)^{T}$ a state on $\HI^{\otimes k}$, the transposed input marginal of the probe, and the collection $\{T_m\}$ is the tester. Conversely, every collection obeying Eq.~(\ref{eq:tester_constraints}) is a physical strategy, reconstructed explicitly from the tester~\cite{Ziman2008,Chiribella2009,Bavaresco2024}. The probe is $\rho_0=\ket{\Phi_\sigma}\!\bra{\Phi_\sigma}$ with
\begin{equation}\label{eq:probe_recon}
    \ket{\Phi_\sigma}=\sum_a\ket{a}\otimes\sigma^{1/2}\ket{a}
\end{equation}
on $\HI^{\otimes k}\otimes\HA^{\otimes k}$, and the measurement is
\begin{equation}\label{eq:povm_recon}
    M_m=(\Id{O}\otimes\sigma^{-1/2})\,W\,T_m\,W^{\dagger}\,
    (\Id{O}\otimes\sigma^{-1/2}),
\end{equation}
where $W$ is the unitary exchanging the two tensor factors, $W(\ket{a}\otimes\ket{o})=\ket{o}\otimes\ket{a}$, and $\sigma^{-1/2}$ is the inverse square root on the support of $\sigma$, the set completed to a POVM by the zero-probability element $\Id{O}\otimes(\Id{}-\Pi)$ with $\Pi$ the support projector of $\sigma$. Inserting this pair into Eq.~(\ref{eq:tester_def}) returns $\{T_m\}$, so the two conditions characterize the physical strategies exactly, the matched ancilla configuration is fully general, and the optimization over testers carries no relaxation.

Substituting $p(m|\tb)=\tr[T_m J_{\tb}]$ into Eq.~(\ref{eq:mse}) makes the risk linear in $\{T_m\}$ for fixed estimates, and the constraints~(\ref{eq:tester_constraints}) are semidefinite and linear, so the minimization over strategies at fixed estimates is a semidefinite program (SDP)~\cite{Vandenberghe1996} solvable to global optimality. Conversely, at fixed tester the optimal estimates are the posterior means in closed form. Alternating the two updates defines a seesaw whose fixed points are candidate optimal strategies~\cite{Bavaresco2024,Andre2026}. We note that each update solves a convex subproblem to global optimality, but the joint problem is bilinear in the tester and the estimates, so a fixed point of the alternation is not guaranteed to be a global optimum of the joint problem. Section~\ref{sec:results} explains how the computed strategies are nevertheless certified, using the NH bound as an external witness. The optimal probe and measurement follow from the converged tester through Eqs.~(\ref{eq:probe_recon}) and~(\ref{eq:povm_recon}). Throughout this work we restrict to the parallel ancilla-assisted class described above.

\section{\label{sec:su2}The $\su$ estimation problem and its symmetry reduction}

The channel is a qubit rotation followed by depolarizing noise,
\begin{equation}\label{eq:su2channel}
    \Lambda_{\tb}=\mathcal{D}_\lambda\circ\,\mathcal{U}_{\tb},
    \qquad
    \mathcal{U}_{\tb}[\rho]=U_{\tb}\,\rho\,U_{\tb}^\dagger,
\end{equation}
with $U_{\tb}=e^{-i\tb\cdot\bm{\sigma}} =\cos r\,\Id{}-i\sin r\,(\nh\cdot\bm{\sigma})$, where $\bm{\sigma}=(\sigma_x,\sigma_y,\sigma_z)$ are the Pauli matrices, $r=\|\tb\|$, $\nh=\tb/r$, and $\mathcal{D}_\lambda[\rho]=(1-\lambda)\rho+\lambda\Id{}/2$ is the qubit depolarizing channel. The prior is the flat density on the ball of radius $r_\mathrm{p}$ in the rotation-vector parametrization,
\begin{align}
\begin{split}
\label{eq:prior}
    p(\tb)&=\frac{3}{4\pi r_\mathrm{p}^{3}}\quad\text{for }\|\tb\|\le r_\mathrm{p},\\
    \qquad p(\tb)&=0\ \text{otherwise},
\end{split}
\end{align}
with equal weights $w_i=1/3$, for which $\mse_\mathrm{prior}=r_\mathrm{p}^2/5$. 

The Choi operator~(\ref{eq:choi}) of a single use is built from the normalized encoded vector
\begin{equation}\label{eq:vdef}
    \ket{v(\tb)}=\frac{1}{\sqrt2}\sum_{i}\ket{i}\otimes
    U_{\tb}\ket{i},
\end{equation}
and a direct computation gives
\begin{equation}\label{eq:choi_su2}
    C(\tb)=2(1-\lambda)\ket{v(\tb)}\!\bra{v(\tb)}
    +\tfrac{\lambda}{2}\,\Id{4},
\end{equation}
with $\tr C(\tb)=d_I=2$, as required of a trace-preserving channel. For $k$ parallel uses the Choi operator is $C(\tb)^{\otimes k}$, the input and output factors collected into $\HI^{\otimes k}\otimes\HO^{\otimes k}$.

\textit{Rotational covariance and representation content.} A rotation $R\in\so$ sends $\tb\to R\tb$ and is implemented on the qubit by the corresponding $\su$ rotation, $U_{R\tb}=V_R\,U_{\tb}\,V_R^\dagger$, with $V_R\in\su$ the representative of $R$. Rotations leave the rest of the problem unchanged. The prior depends only on $r$, the equal-weight cost $\|\hat{\tb}-\tb\|^2/3$ only on $\|\hat{\tb}-\tb\|$, and the noise obeys $\mathcal{D}_\lambda[V\rho V^\dagger] =V\mathcal{D}_\lambda[\rho]V^\dagger$, so the estimation problem has the full rotational symmetry. Since
\begin{equation}\label{eq:v_transform}
\begin{split}
    \ket{v(R\tb)}&=\frac{1}{\sqrt2}\sum_i\ket{i}\otimes
    V_RU_{\tb}V_R^\dagger\ket{i}\\
    &=(\bar V_R\otimes V_R)\ket{v(\tb)},
\end{split}
\end{equation}
the $k$-use Choi operator transforms by the conjugation representation
\begin{equation}\label{eq:Wrep}
    \Wrep_R=(\bar V_R\otimes V_R)^{\otimes k},
    \qquad
    J_{R\tb}=\Wrep_R\,J_{\tb}\,\Wrep_R^{\dagger}.
\end{equation}
The single-use Choi space splits into a one-dimensional spin-$0$ subspace spanned by $\ket{0_s}=\ket{v(\bm 0)}$ and a three-dimensional spin-$1$ subspace spanned by $\ket{a_s}=\bigl(\sum_i\ket{i}\otimes\sigma_a\ket{i}\bigr)/\sqrt2$, so that
\begin{equation}\label{eq:v_decomp}
    \ket{v(\tb)}=\cos r\,\ket{0_s}
    -i\sin r\,n_a\ket{a_s}.
\end{equation}
On $k$ copies the Choi space carries $(0\oplus1)^{\otimes k}$, with spin-$J$ multiplicities $m_J^{(k)}$ fixed by the Clebsch--Gordan series.

\textit{Seed reduction of the strategy and NH programs.} Covariant problems admit covariant optima~\cite{Holevo2011book}, here strategies obeying $T(R\hat{\tb})=\Wrep_R\,T(\hat{\tb})\,\Wrep_R^{\dagger}$ with the estimate as the outcome label, $\hat{\tb}=\hat r\,\nh$, and the completeness condition $\sum_{\hat r}\int(d\nh/4\pi)\,T(\hat r,\nh)=\sigma\otimes\Id{O}$, where $\hat r$ runs over a grid of $n_r$ values on $[0,r_\mathrm{p}]$ while $\nh$ ranges over the whole sphere, so the estimates are discretized only in magnitude, and $\sigma$ is taken rotation invariant. We define the \emph{seed} $\seed(\hat r)$ as the tester element whose estimate has magnitude $\hat r$ and points along the reference direction $\hat z$. Covariance under any rotation $R(\nh)$ with $R(\nh)\hat z=\nh$ then generates every other element from the seed,
\begin{equation}\label{eq:seed_family}
\begin{gathered}
    \seed(\hat r):=T(\hat r,\hat z)\geq0,\\
    T(\hat r,\nh)=\Wrep_{R(\nh)}\,\seed(\hat r)\,
    \Wrep_{R(\nh)}^{\dagger},
\end{gathered}
\end{equation}
so the tester is specified by the $n_r$ seeds, one per radial grid point. The rotation $R(\nh)$ is fixed only up to replacing it by $R(\nh)R_z(\phi)$, with $R_z(\phi)$ the rotation by $\phi$ about $\hat z$. These rotations leave $\hat z$ fixed, so covariance gives $\Wrep_{R_z(\phi)}\seed(\hat r)\Wrep_{R_z(\phi)}^{\dagger}=\seed(\hat r)$, and this invariance makes Eq.~(\ref{eq:seed_family}) independent of the choice. Conversely, every positive operator with this invariance generates a covariant family through Eq.~(\ref{eq:seed_family}). Writing $\Wrep_{R_z(\phi)}=e^{-i\phi J_z}$, the invariance says that the seed commutes with the generator $J_z$, so it is block diagonal in the weight $m$, the eigenvalue of $J_z$,
\begin{equation}\label{eq:seed_blocks}
    \seed(\hat r)=\bigoplus_{m}\seed_m(\hat r),
    \qquad \seed_m(\hat r)\geq 0 ,
\end{equation}
with block dimensions $d_m=\sum_{J\geq|m|}m_J^{(k)}=\binom{2k}{k+m}$. The weight blocks $\seed_m(\hat r)$, with $|m|\leq k$, are therefore the variables of the strategy SDP, together with $\sigma$ when the probe is optimized. The seed follows from them through Eq.~(\ref{eq:seed_blocks}), the tester through Eq.~(\ref{eq:seed_family}), and the probe and the POVM through Eqs.~(\ref{eq:probe_recon}) and~(\ref{eq:povm_recon}). Positivity splits into one condition per block, each of size at most $d_0=\binom{2k}{k}$ rather than $4^{k}$, and the blocks are coupled only through the completeness condition. Averaging the covariant family over directions is the Haar twirl~\cite{Bartlett2007,Watrous2018}, so both sides of the completeness condition are invariant under every $\Wrep_R$. Explicitly, $\int dR\,\Wrep_R\,A\,\Wrep_R^\dagger$ with $dR$ the normalized Haar measure projects any operator $A$ onto the commutant of the representation $\Wrep$, and the completeness condition~(\ref{eq:tester_constraints}) equates two operators already in that commutant. By the Schur lemma such operators carry only $\sum_J(m_J^{(k)})^2=C_{2k}=\binom{4k}{2k}/(2k+1)$ independent components, far fewer than the $16^{k}$ of a general operator, so completeness reduces to $C_{2k}$ linear equations relating the seed blocks to $\sigma$, and this reduction carries the optimization to values of $k$ out of reach for the direct programs. Rotating every direction to $\hat z$ inside Eq.~(\ref{eq:mse}) collapses the objective to
\begin{equation}\label{eq:obj_blocks}
    \mse=\sum_{\hat r,m}\tfrac13\,
    \tr\!\bigl[\bigl(\hat r^{\,2} A_{0,m}-2\hat r\,A_{z,m}
    +A_{2,m}\bigr)\,\seed_m(\hat r)\bigr],
\end{equation}
where $A_{0,m}=P_mK_0P_m$, $A_{z,m}=P_mK_zP_m$, and $A_{2,m}=P_mK_2P_m$ are the weight-$m$ blocks of the three Choi moments
\begin{align}
    K_0&=\int d\tb\,p(\tb)\,J_{\tb}\,,\label{eq:K0}\\
    K_z&=\int d\tb\,p(\tb)\,\theta_z\,J_{\tb}\,,\label{eq:Kz}\\
    K_2&=\int d\tb\,p(\tb)\,r^2\,J_{\tb}\,,\label{eq:K2}
\end{align}
evaluated exactly, with $P_m$ the projector onto the weight-$m$ eigenspace of $J_z$. $K_0$ and $K_2$ lie in the commutant by the invariance of the prior, and $K_z$ is weight diagonal because $\theta_z$ is invariant under rotations about $\hat z$, so the diagonal blocks carry the entire content of all three moments. Minimizing Eq.~(\ref{eq:obj_blocks}) over the blocks $\seed_m(\hat r)\geq0$ under the reduced completeness equations, jointly with the invariant $\sigma$, is the strategy SDP, alternated with the posterior-mean update in the seesaw of Sec.~\ref{sec:testers}.

The same reduction applies to the NH program~(\ref{eq:nh_sdp}), whose variables $\Lop$ and $X$ act on $\HO^{\otimes k}\otimes\HA^{\otimes k}$ of dimension $D=4^k$. Both probe classes have a rotation-invariant input marginal, so the output family is covariant, $\rho(R\tb)=G_R\,\rho(\tb)\,G_R^{\dagger}$ with $G_R=V_R^{\otimes k}\otimes\bar V_R^{\otimes k}$, since the depolarizing noise commutes with unitary conjugation and the channel does not act on the ancilla. Changing variables $\tb\to R\tb$ in the moments (\ref{eq:moment0}) and (\ref{eq:moment1}) then gives $G_R\Gz G_R^{\dagger}=\Gz$ and $G_R\Go{i}G_R^{\dagger}=\sum_j R_{ji}\Go{j}$, a rotational scalar and a rotational vector. This symmetry of the data carries over to the NH program, and we use it to reduce the program in three steps. We let each rotation act on the NH variables, check that this action preserves both the objective and the constraint, and average over rotations to obtain an optimal pair that is itself rotation invariant and therefore block diagonal by the Schur lemma. The action on the variables mirrors the one on the moments, which are conjugated by $G_R$ and, for the first moment, also rotated in the parameter label. Rotating every parameter index by $R$ and conjugating every operator entry by $G_R$ defines the \emph{transported} pair
\begin{equation}\label{eq:transport}
\begin{gathered}
    X_i^R=\sum_{i'}R_{ii'}\,G_RX_{i'}G_R^{\dagger},\\
    \Lop_{ij}^R=\sum_{i',j'}R_{ii'}R_{jj'}\,G_R\Lop_{i'j'}G_R^{\dagger}.
\end{gathered}
\end{equation}
The objective of Eq.~(\ref{eq:nh_sdp}) is unchanged, because $\sum_iR_{ij}R_{il}=\delta_{jl}$ cancels the vector rotation of $\Go{i}$ against the index rotation of $X_i$, a cancellation that requires the equal weights. Feasibility is also unchanged. By the Schur complement~\cite{Boyd2004,Zhang2005} the constraint $\Lop\geq XX^{\intercal}$ is the positivity of the block matrix
\begin{equation}\label{eq:nh_block}
    \mathbb{M}=\begin{pmatrix}\Lop & X\\ X^{\intercal} & \Id{}
    \end{pmatrix}
\end{equation}
on $(\mathbb{C}^3\oplus\mathbb{C})\otimes\HO^{\otimes k}\otimes \HA^{\otimes k}$, the transport~(\ref{eq:transport}) is conjugation by the unitary $U_R=(R\oplus1)\otimes G_R$, and $U_R$ preserves positivity, the identity corner, and the symmetry $\Lop_{ij}=\Lop_{ji}$. Averaging any feasible pair over rotations therefore yields an invariant pair of equal objective, so restricting to invariant $\mathbb{M}$ leaves $\mse_\mathrm{NH}$ unchanged~\cite{Bachoc2012invariant}. That restriction is the only one taken, the subsequent change of basis loses nothing. Since $R\oplus1$ carries spin $0\oplus1$, the content of one Choi factor, $U_R$ is equivalent to $(\bar V_R\otimes V_R)^{\otimes(k+1)}$, and by the Schur lemma an invariant $\mathbb{M}$ is block diagonal in the spin basis of this representation, one positive block of size $m_J^{(k+1)}$ per spin $J$, with $\sum_J\bigl(m_J^{(k+1)}\bigr)^2$ real parameters. The identity corner and the symmetry of $\Lop$ become linear conditions on the blocks, and the solution rotates back to $\Lop$ and $X$. The direct and the reduced programs, solved independently at $k=1$ and $2$, agree to $1.4\times10^{-8}$ over all noise strengths and both probe classes.

\begin{figure}[t]
\includegraphics[width=\columnwidth]{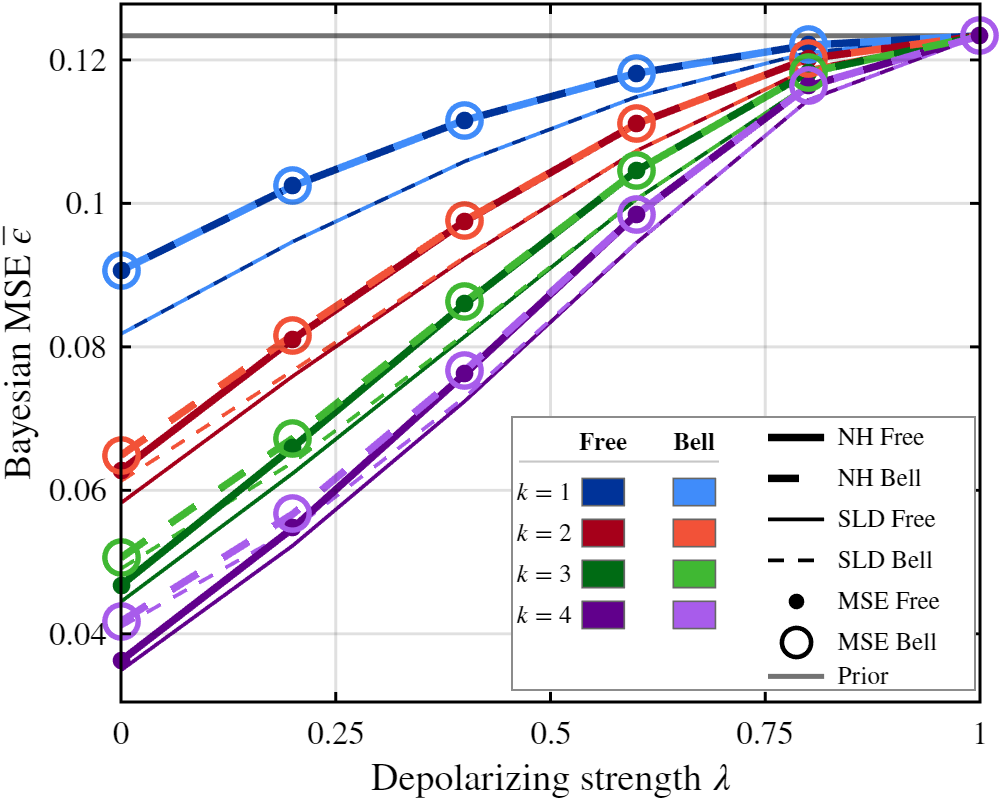}
\caption{\label{fig:bounds}Bayes risk and bounds for the $\su$ channel as a function of the depolarizing strength $\lambda$. Each number of channel uses $k$ is one hue, dark solid lines for the optimized probe and bright dashed lines of the same hue for the tensored Bell probe. Thick lines are the NH bound and thin lines the SLD bound. Markers are the achieved risk $\mse^\star$, filled dots for the optimized probe and open circles for the Bell probe. Markers lying on the thick lines show attainment of the NH bound, and the offset to the thin lines is the incompatibility cost. The two achieved-risk curves of each hue meet as $\lambda$ grows, past which the tensored Bell probe is optimal. The gray line is the prior risk $r_\mathrm{p}^2/5$, approached by every curve as $\lambda\to1$. All curves and markers are evaluated from the exact moments. Numerical parameters: the radial estimate grid has $n_r=21$ points uniformly spaced on $[0,r_\mathrm{p}]$, the parameter ball is sampled on a lattice of $14147$ points, a $31^{3}$ cubic lattice on $[-r_\mathrm{p},r_\mathrm{p}]^{3}$ restricted to the ball with equal weights, the depolarizing strength is sampled at 6 values of $\lambda$, and all semidefinite programs are solved with MOSEK 11.1.8 through YALMIP to its default primal-dual tolerance of $10^{-8}$.}
\end{figure}

\section{\label{sec:results}Results}

All results take the ball radius $r_\mathrm{p}=\pi/4$, for which $\mse_\mathrm{prior}=\pi^2/80\approx0.1234$, and two probe classes, which differ only in how the transposed input marginal $\sigma$ of Eq.~(\ref{eq:probe_recon}) is treated. For the tensored Bell probe $\ket{\Phi^+}^{\otimes k}$, $\sigma=\Id{}/2^{k}$ is held fixed. For the optimized probe, $\sigma$ is a variable of the strategy SDP, constrained only to be a state invariant under the collective rotation $V_R^{\otimes k}$ and optimized jointly with the seed blocks, so this probe is a numerical output of the SDP, reconstructed from the returned $\sigma$ through Eq.~(\ref{eq:probe_recon}). The tester seesaw is optimized over a discretized problem, the radial estimate grid of Sec.~\ref{sec:su2} with $n_r=21$ points and a lattice over the parameter ball of $14147$ points, the $31^{3}$
cubic lattice of Fig.~\ref{fig:bounds}. The moments $\Gz$ and $\Go{i}$ entering $\mse_\mathrm{SLD}$, $\mse_\mathrm{NH}$, and the evaluation of $\mse^\star$ are computed exactly. By Eqs.~(\ref{eq:choi_su2}) and~(\ref{eq:v_decomp}) their integrands are polynomials of angular degree at most $2k$, or $2k+1$ for the first moments, so a product rule with $k+1$ Gauss--Legendre nodes in $\cos\theta_p$ and $2k+2$ equispaced nodes in $\phi$ reproduces every angular integral with zero error, the radial integrals are evaluated by Gauss--Legendre quadrature to machine precision, and the scalar second moment is $r_\mathrm{p}^2/5$ in closed form.

\begin{figure}[t]
\includegraphics[width=\columnwidth]
{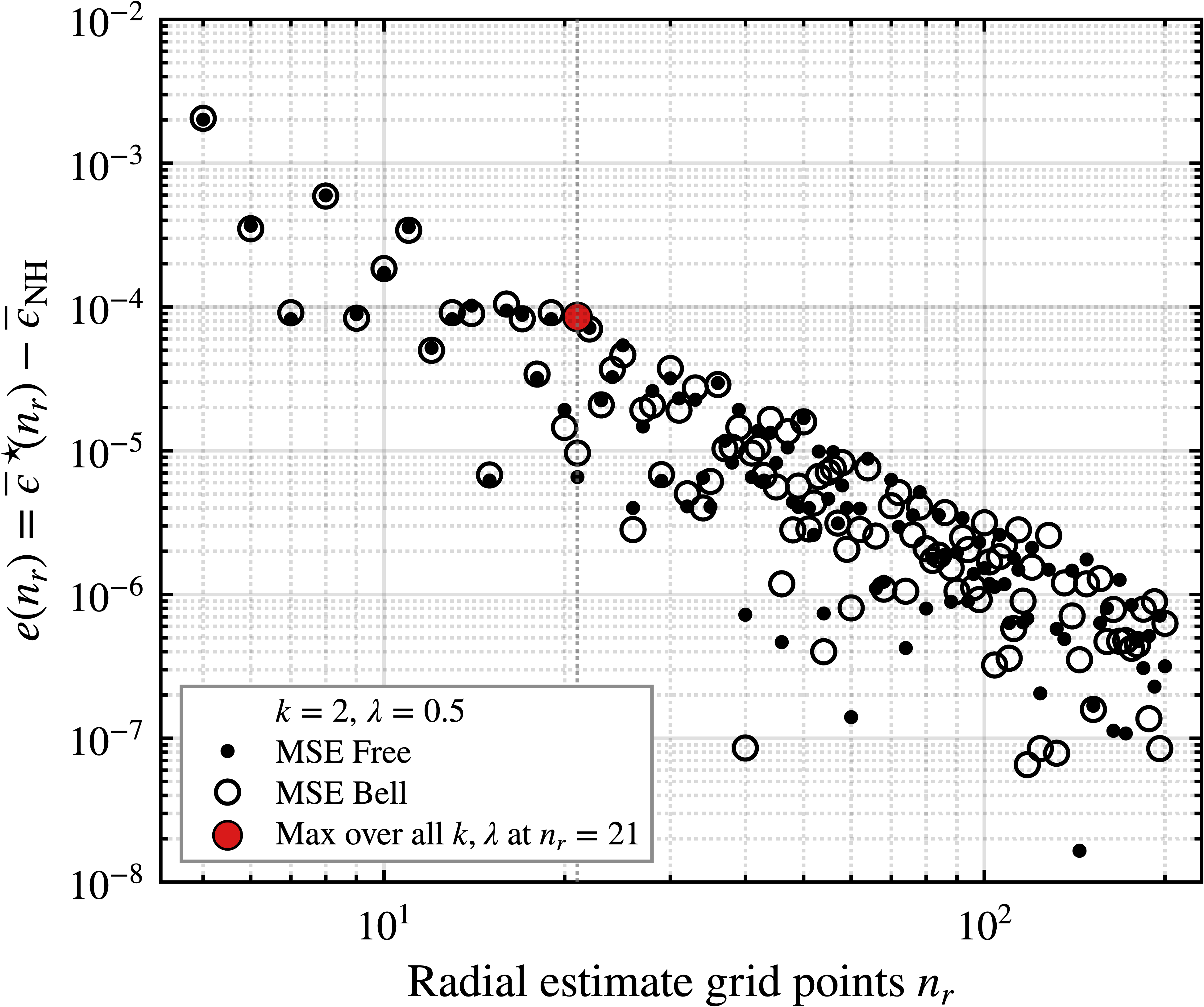}
\caption{\label{fig:sweep}Deviation $e(n_r)=\mse^\star(n_r)-\mse_\mathrm{NH}$ of the achieved risk from the NH bound as a function of the number $n_r$ of radial estimate-grid points, at $k=2$ and $\lambda=0.5$. Filled black dots are the optimized probe with $\sigma$ held fixed at its optimum for $n_r=21$, and open circles the tensored Bell probe. The red dot marks $8.5\times10^{-5}$, the largest deviation over all $k$ and $\lambda$ in Fig.~\ref{fig:bounds}, at the grid size $n_r=21$ used there, shown by the dotted line. The scatter reflects where the uniform grid points fall relative to the optimal estimate radii. All other numerical parameters are as in Fig.~\ref{fig:bounds}.}
\end{figure}

Figure~\ref{fig:bounds} shows $\mse^\star$, $\mse_\mathrm{SLD}$, and $\mse_\mathrm{NH}$ for both probe classes and $k=1$ to $4$. For every $k$ and $\lambda$ the achieved risk exceeds the NH bound by at most $8.5\times10^{-5}$, below one part in $10^{3}$ of the bound. This small deviation is expected, since the measurement and its estimates are optimized for the discretized problem rather than for the exact moments, so $\mse^\star$ is the exactly evaluated risk of a slightly suboptimal strategy and upper bounds the exact optimum of Eq.~(\ref{eq:mse}). 
To confirm that the deviation is of discretization origin rather than a genuine gap, we have studied its behavior under refinement of the estimate grid. Figure~\ref{fig:sweep} shows $e(n_r)=\mse^\star(n_r)-\mse_\mathrm{NH}$ at $k=2$ and $\lambda=0.5$ for 106 values of $n_r$ between 5 and 200, for the tensored Bell probe and for the optimized probe with $\sigma$ held fixed at its optimum for $n_r=21$, so that only the estimate grid varies. All other numerical parameters are as in Fig.~\ref{fig:bounds}. At $n_r=21$, the grid size of Fig.~\ref{fig:bounds}, the deviation is $9.6\times10^{-6}$ for the Bell probe and $6.6\times10^{-6}$ for the optimized probe, well below the largest deviation found there, $8.5\times10^{-5}$, which is marked in Fig.~\ref{fig:sweep}. The deviation scatters from one value of $n_r$ to the next, because it depends on where the uniform grid points fall relative to the optimal estimate radii, but its largest values decrease steadily, from $2\times10^{-3}$ at $n_r=5$ to below $10^{-6}$ for every $n_r\geq170$, with no sign of saturation. A genuine gap would saturate at a finite value under refinement, which is not what is observed. The deviation is also insensitive to the solver tolerance, changing by less than $0.2\%$ between primal-dual tolerances of $10^{-8}$ and $10^{-10}$ at $n_r=57$, and to the moment lattice, changing by less than $1\%$ when the lattice is refined from $31$ to $51$ points per axis at $n_r=113$. The residual is therefore set by the estimate grid rather than by the interior-point solve or the lattice. Whether the Bayesian NH bound can be attained is an open problem~\cite{Suzuki2024}, and we do not prove attainment. The near coincidence of an achievable risk with the lower bound is numerical evidence that the bound is attained for this channel family at every accessible $k$, and it also certifies the computed strategies as globally optimal to the same precision. The certificate deserves a word, since the seesaw of Sec.~\ref{sec:testers} is an alternating optimization of a jointly bilinear problem and carries no intrinsic guarantee of global optimality. The guarantee comes from outside the algorithm. The quantity $\mse_\mathrm{NH}$ lower bounds the Bayes risk of every strategy in the class, and it is computed by a single convex program whose optimum is certified by the duality gap returned by the solver, with no seesaw involved. Whenever the exactly evaluated risk of a converged seesaw strategy meets that lower bound to within a stated tolerance, no strategy in the class can do better by more than that tolerance, and the strategy is optimal to that precision irrespective of how it was found. The optimality we claim is thus conditional on the observed coincidence and is a statement about the certificate, not a property of the seesaw. The NH bound is strictly above the SLD bound at every $\lambda<1$, so the SLD bound is unattained and, by the SLD attainability condition, the Bayesian SLD operators of this family never commute. The attainable error can exceed the SLD bound by at most a factor of two~\cite{Albarelli2026}. For the optimized probe and the noiseless channel the observed ratio $\mse_\mathrm{NH}/\mse_\mathrm{SLD}$ falls from $1.108$ at $k=1$ to $1.038$ at $k=4$, closing part of the gap to the SLD bound as the number of uses grows, and we make no scaling claim from this range. As the depolarizing strength grows the achieved risks of the two probe classes merge, visible in Fig.~\ref{fig:bounds} as the meeting of each pair of curves. There the optimized probe itself collapses to the tensored Bell state.

\section{\label{sec:discussion}Discussion}

We presented a strategy that jointly optimizes the probe, the measurement, and the estimator for $k$ uses of the depolarized $\su$ rotation channel, by reducing the rotationally invariant problem to small convex programs for both the strategy search and the NH bound. The optimized strategy reaches the Bayesian NH bound within the numerical precision of the calculation at every computed $k$ and noise strength, and as the noise grows the optimal probe collapses to the tensored Bell state.

Several directions remain open. The restriction to parallel strategies is without loss for the noiseless channel, where memory-assisted schemes offer no advantage for independent unitaries~\cite{ChiribellaPRL180501}, but sequential strategies can outperform parallel ones under noise, as found for the $\su$ rotation under amplitude damping~\cite{Andre2026}, and extending the reduction to that class is open. If a condition for the attainability of the Bayesian NH bound is proved~\cite{Suzuki2024}, this channel family is a natural test case. Higher-dimensional channels are out of reach of the present programs, and whether the observed approach of the NH bound toward the SLD bound continues at larger $k$ is an open question.

\begin{acknowledgments}
C.N.G. and L.B. acknowledge the award ARO W911NF-24-1-0080. L.B. acknowledges support from the Alfred P. Sloan Foundation (Grant Award No. G-2020-12684) through the University of Arizona Sloan Indigenous Graduate Partnership and from the Gates Millennium Scholars Program, funded by the Bill and Melinda Gates Foundation.
\end{acknowledgments}
\bibliography{references}

@article{Rubio2020,
  title={Bayesian multiparameter quantum metrology with limited data},
  author={Rubio, Jes{\'u}s and Dunningham, Jacob},
  journal={Physical Review A},
  volume={101},
  number={3},
  pages={032114},
  year={2020},
  publisher={APS}
}

@article{Andre2026,
  title={Strategy optimization for Bayesian quantum parameter estimation with finite copies: Adaptive greedy, parallel, sequential, and general strategies},
  author={L. Andr{\'e}, Erik and Bavaresco, Jessica and Mehboudi, Mohammad},
  journal={Quantum Science and Technology},
  year={2026}
}

@article{Albarelli2026,
  title={Measurement incompatibility in Bayesian multiparameter quantum estimation},
  author={Albarelli, Francesco and Branford, Dominic and Rubio, Jes{\'u}s},
  journal={arXiv preprint arXiv:2511.16645},
  year={2025}
}

@article{Chiribella2009,
  title={Theoretical framework for quantum networks},
  author={Chiribella, Giulio and D’Ariano, Giacomo Mauro and Perinotti, Paolo},
  journal={Physical Review A—Atomic, Molecular, and Optical Physics},
  volume={80},
  number={2},
  pages={022339},
  year={2009},
  publisher={APS}
}

@article{Braunstein1994,
  title={Statistical distance and the geometry of quantum states},
  author={Braunstein, Samuel L and Caves, Carlton M},
  journal={Physical Review Letters},
  volume={72},
  number={22},
  pages={3439},
  year={1994},
  publisher={APS}
}

@article{Helstrom1974,
  title={Noncommuting observables in quantum detection and estimation theory},
  author={Helstrom, C and Kennedy, R},
  journal={IEEE Transactions on Information Theory},
  volume={20},
  number={1},
  pages={16--24},
  year={1974},
  publisher={IEEE}
}

@article{Vandenberghe1996,
  title={Semidefinite programming},
  author={Vandenberghe, Lieven and Boyd, Stephen},
  journal={SIAM review},
  volume={38},
  number={1},
  pages={49--95},
  year={1996},
  publisher={SIAM}
}

@incollection{Nagaoka1989,
  title={On the parameter estimation problem for quantum statistical models},
  author={Nagaoka, Hiroshi},
  booktitle={Asymptotic Theory of Quantum Statistical Inference: Selected Papers},
  pages={125--132},
  year={2005},
  publisher={World Scientific}
}

@article{Suzuki2024,
  title={Bayesian Nagaoka-Hayashi bound for multiparameter quantum-state estimation problem},
  author={Suzuki, Jun},
  journal={IEICE Transactions on Fundamentals of Electronics, Communications and Computer Sciences},
  volume={107},
  number={3},
  pages={510--518},
  year={2024},
  publisher={The Institute of Electronics, Information and Communication Engineers}
}

@article{Bavaresco2024,
  title={Designing optimal protocols in Bayesian quantum parameter estimation with higher-order operations},
  author={Bavaresco, Jessica and Lipka-Bartosik, Patryk and Sekatski, Pavel and Mehboudi, Mohammad},
  journal={Physical Review Research},
  volume={6},
  number={2},
  pages={023305},
  year={2024},
  publisher={APS}
}

@article{HayashiOuyang2023,
  title={Tight Cram{\'e}r-Rao type bounds for multiparameter quantum metrology through conic programming},
  author={Hayashi, Masahito and Ouyang, Yingkai},
  journal={Quantum},
  volume={7},
  pages={1094},
  year={2023},
  publisher={Verein zur F{\"o}rderung des Open Access Publizierens in den Quantenwissenschaften}
}

@article{HayashiOuyang2024probe,
  title={Finding the optimal probe state for multiparameter quantum metrology using conic programming},
  author={Hayashi, Masahito and Ouyang, Yingkai},
  journal={npj Quantum Information},
  volume={10},
  number={1},
  pages={111},
  year={2024},
  publisher={Nature Publishing Group UK London}
}

@article{Personick1971,
  title={Application of quantum estimation theory to analog communication over quantum channels},
  author={Personick, S},
  journal={IEEE Transactions on Information Theory},
  volume={17},
  number={3},
  pages={240--246},
  year={1971},
  publisher={IEEE}
}

@article{YuenLax1973,
  title={Multiple-parameter quantum estimation and measurement of nonselfadjoint observables},
  author={Yuen, Horace and Lax, Melvin},
  journal={IEEE Transactions on Information Theory},
  volume={19},
  number={6},
  pages={740--750},
  year={1973},
  publisher={IEEE}
}

@article{Holevo1973,
  title={Statistical decision theory for quantum systems},
  author={Holevo, Alexander S},
  journal={Journal of multivariate analysis},
  volume={3},
  number={4},
  pages={337--394},
  year={1973},
  publisher={Elsevier}
}

@article{Demkowicz2020review,
  title={Multi-parameter estimation beyond quantum Fisher information},
  author={Demkowicz-Dobrza{\'n}ski, Rafa{\l} and G{\'o}recki, Wojciech and Gu{\c{t}}{\u{a}}, M{\u{a}}d{\u{a}}lin},
  journal={Journal of Physics A: Mathematical and Theoretical},
  volume={53},
  number={36},
  pages={363001},
  year={2020},
  publisher={IOP Publishing}
}

@article{Demkowicz2011phase,
  title={Optimal phase estimation with arbitrary a priori knowledge},
  author={Demkowicz-Dobrza{\'n}ski, Rafa{\l}},
  journal={Physical Review A—Atomic, Molecular, and Optical Physics},
  volume={83},
  number={6},
  pages={061802},
  year={2011},
  publisher={APS}
}

@article{Chiribella2005group,
  title={Optimal estimation of group transformations using entanglement},
  author={Chiribella, Giulio and D’ariano, GM and Sacchi, Massimiliano Federico},
  journal={Physical Review A—Atomic, Molecular, and Optical Physics},
  volume={72},
  number={4},
  pages={042338},
  year={2005},
  publisher={APS}
}

@article{Chiribella2008architecture,
  title={Quantum circuit architecture},
  author={Chiribella, Giulio and D’Ariano, G Mauro and Perinotti, Paolo},
  journal={Physical review letters},
  volume={101},
  number={6},
  pages={060401},
  year={2008},
  publisher={APS}
}

@article{Acin2001dynamics,
  title={Optimal estimation of quantum dynamics},
  author={Ac{\'\i}n, Antonio and Jan{\'e}, E and Vidal, Guifr{\'e}},
  journal={Physical Review A},
  volume={64},
  number={5},
  pages={050302},
  year={2001},
  publisher={APS}
}

@article{Conlon2021,
  title={Efficient computation of the Nagaoka--Hayashi bound for multiparameter estimation with separable measurements},
  author={Conlon, Lorc{\'a}n O and Suzuki, Jun and Lam, Ping Koy and Assad, Syed M},
  journal={npj Quantum Information},
  volume={7},
  number={1},
  pages={110},
  year={2021},
  publisher={Nature Publishing Group UK London}
}

@article{Tsang2019thrice,
  title={The holevo cram$\backslash$'er-rao bound is at most thrice the helstrom version},
  author={Tsang, Mankei},
  journal={arXiv preprint arXiv:1911.08359},
  year={2019}
}

@article{Tsang2020GL,
  title={Physics-inspired forms of the Bayesian Cram{\'e}r-Rao bound},
  author={Tsang, Mankei},
  journal={Physical Review A},
  volume={102},
  number={6},
  pages={062217},
  year={2020},
  publisher={APS}
}

@article{ChiribellaPRL180501,
  title={Memory effects in quantum channel discrimination},
  author={Chiribella, Giulio and D’Ariano, Giacomo M and Perinotti, Paolo},
  journal={Physical review letters},
  volume={101},
  number={18},
  pages={180501},
  year={2008},
  publisher={APS}
}

@article{Bagan2004,
  title={Entanglement-assisted alignment of reference frames using a dense covariant coding},
  author={Bagan, Emilio and Baig, Mariano and Munoz-Tapia, Ramon},
  journal={Physical Review A—Atomic, Molecular, and Optical Physics},
  volume={69},
  number={5},
  pages={050303},
  year={2004},
  publisher={APS}
}

@book{Holevo2011book,
  title={Probabilistic and statistical aspects of quantum theory},
  author={Holevo, Alexander S},
  volume={1},
  year={2011},
  publisher={Springer Science \& Business Media}
}

@article{Helstrom1967,
  title={Minimum mean-squared error of estimates in quantum statistics},
  author={Helstrom, Carl W},
  journal={Physics letters A},
  volume={25},
  number={2},
  pages={101--102},
  year={1967},
  publisher={Elsevier}
}

@article{Giovannetti2011advances,
  title={Advances in quantum metrology},
  author={Giovannetti, Vittorio and Lloyd, Seth and Maccone, Lorenzo},
  journal={Nature photonics},
  volume={5},
  number={4},
  pages={222--229},
  year={2011},
  publisher={Nature Publishing Group UK London}
}

@article{Ragy2016compatibility,
  title={Compatibility in multiparameter quantum metrology},
  author={Ragy, Sammy and Jarzyna, Marcin and Demkowicz-Dobrza{\'n}ski, Rafa{\l}},
  journal={Physical Review A},
  volume={94},
  number={5},
  pages={052108},
  year={2016},
  publisher={APS}
}

@article{Kahn2009local,
  title={Local asymptotic normality for finite dimensional quantum systems},
  author={Kahn, Jonas and Gu{\c{t}}{\u{a}}, M{\u{a}}d{\u{a}}lin},
  journal={Communications in Mathematical Physics},
  volume={289},
  number={2},
  pages={597--652},
  year={2009},
  publisher={Springer}
}

@article{Yamagata2013qlan,
   title={Quantum local asymptotic normality based on a new quantum likelihood ratio},
   volume={41},
   ISSN={0090-5364},
   url={http://dx.doi.org/10.1214/13-AOS1147},
   DOI={10.1214/13-aos1147},
   number={4},
   journal={The Annals of Statistics},
   publisher={Institute of Mathematical Statistics},
   author={Yamagata, Koichi and Fujiwara, Akio and Gill, Richard D.},
   year={2013},
   month=Aug }

@article{Yang2019attaining,
  title={Attaining the ultimate precision limit in quantum state estimation},
  author={Yang, Yuxiang and Chiribella, Giulio and Hayashi, Masahito},
  journal={Communications in Mathematical Physics},
  volume={368},
  number={1},
  pages={223--293},
  year={2019},
  publisher={Springer}
}

@incollection{Nagaoka1991,
  title={A generalization of the simultaneous diagonalization of Hermitian matrices and its relation to quantum estimation theory},
  author={Nagaoka, Hiroshi},
  booktitle={Asymptotic Theory of Quantum Statistical Inference: Selected Papers},
  pages={133--149},
  year={2005},
  publisher={World Scientific}
}

@article{Hayashi1999,
  title={On simultaneous measurement of noncommutative observables},
  author={Hayashi, Masahito},
  journal={Development of infinite-dimensional non-commutative anaysis},
  pages={96--188},
  year={1999}
}

@article{TsangZivZakai2012,
  title={Ziv-Zakai error bounds for quantum parameter estimation},
  author={Tsang, Mankei},
  journal={Physical review letters},
  volume={108},
  number={23},
  pages={230401},
  year={2012},
  publisher={APS}
}

@article{LuTsang2016,
  title={Quantum Weiss-Weinstein bounds for quantum metrology},
  author={Lu, Xiao-Ming and Tsang, Mankei},
  journal={Quantum Science and Technology},
  volume={1},
  number={1},
  pages={015002},
  year={2016},
  publisher={IOP Publishing}
}

@article{SidhuKok2020,
  title={Geometric perspective on quantum parameter estimation},
  author={Sidhu, Jasminder S and Kok, Pieter},
  journal={AVS Quantum Science},
  volume={2},
  number={1},
  year={2020},
  publisher={AIP Publishing}
}

@inproceedings{Gutoski2007,
  title={Toward a general theory of quantum games},
  author={Gutoski, Gus and Watrous, John},
  booktitle={Proceedings of the thirty-ninth annual ACM symposium on Theory of computing},
  pages={565--574},
  year={2007}
}

@article{Ziman2008,
  title={Process positive-operator-valued measure: A mathematical framework for the description of process tomography experiments},
  author={Ziman, M{\'a}rio},
  journal={Physical Review A—Atomic, Molecular, and Optical Physics},
  volume={77},
  number={6},
  pages={062112},
  year={2008},
  publisher={APS}
}

@inbook{Bachoc2012invariant,
   title={Invariant Semidefinite Programs},
   ISBN={9781461407690},
   ISSN={2214-7934},
   url={http://dx.doi.org/10.1007/978-1-4614-0769-0_9},
   DOI={10.1007/978-1-4614-0769-0_9},
   booktitle={Handbook on Semidefinite, Conic and Polynomial Optimization},
   publisher={Springer US},
   author={Bachoc, Christine and Gijswijt, Dion C. and Schrijver, Alexander and Vallentin, Frank},
   year={2011},
   month={Sep}, 
   pages={219269} }

@article{Bartlett2007,
  author  = {Bartlett, Stephen D. and Rudolph, Terry and Spekkens, Robert W.},
  title   = {Reference frames, superselection rules, and quantum information},
  journal = {Rev. Mod. Phys.},
  volume  = {79},
  issue   = {2},
  pages   = {555--609},
  year    = {2007},
  doi     = {10.1103/RevModPhys.79.555}
}

@book{Watrous2018,
  author    = {Watrous, John},
  title     = {The Theory of Quantum Information},
  publisher = {Cambridge University Press},
  address   = {Cambridge},
  year      = {2018},
  doi       = {10.1017/9781316848142}
}

@book{Boyd2004,
  author    = {Boyd, Stephen and Vandenberghe, Lieven},
  title     = {Convex Optimization},
  publisher = {Cambridge University Press},
  address   = {Cambridge},
  year      = {2004},
  note      = {Schur complement: Appendix A.5.5},
  doi       = {10.1017/CBO9780511804441}
}

@book{Zhang2005,
  editor    = {Zhang, Fuzhen},
  title     = {The Schur Complement and Its Applications},
  series    = {Numerical Methods and Algorithms},
  volume    = {4},
  publisher = {Springer},
  address   = {New York},
  year      = {2005},
  doi       = {10.1007/b105056}
}

\end{document}